\documentclass[twoside,twocolumn,english,12pt]{IEEEtran}
\usepackage[T1]{fontenc}

\makeatletter

 \newenvironment{lyxcode}
   {\begin{list}{}{
     \setlength{\rightmargin}{\leftmargin}
     \setlength{\listparindent}{0pt}
     \raggedright
     \setlength{\itemsep}{0pt}
     \setlength{\parsep}{0pt}
     \normalfont\ttfamily}%
    \item[]}
   {\end{list}}

\usepackage{babel}
\makeatother
\begin{document}

\title{A Fast, Vectorizable Algorithm for Producing Single-Precision Sine-Cosine
Pairs }

\author{Marcus H. Mendenhall\thanks{Marcus Mendenhall is with the Vanderbilt University W. M. Keck Free Electron Laser Center and the Department of Physics and Astronomy.  Research Supported by the United States DOD MFEL program under grant FA9550-04-1-0045}}

\maketitle
\begin{abstract}
This paper presents an algorithm for computing Sine-Cosine pairs to
modest accuracy, but in a manner which contains no conditional tests
or branching, making it highly amenable to vectorization. An exemplary
implementation for PowerPC AltiVec processors is included, but the
algorithm should be easily portable to other achitectures, such as
Intel SSE.
\end{abstract}

\markboth{}{}

\section{Introduction}

\PARstart{T}{he} need for sine-cosine pairs arises often in numerical
computing and digital signal processing. For many of the applications,
the functions are needed at uniformly spaced values of the argument,
and can be generated by efficient and accurate recursion relations. 

For some applications, though, one needs to compute trigonometric
functions in random order, or for non-uniformly spaced values. If
a large number of such evaluations are needed, it would be convenient
to be able to take advantage of the extremely fast vector processing
units included on many modern desktop computer families. Unfortunately,
most common algorithms require that one reduce the angle modulo $\pi/2$
onto the range $[-\pi/4,\pi/4)$ (or sometimes mod $\pi$ onto $[-\pi/2,\pi/2)$)
\cite{AandS}, then to compute the function using some reasonable
power series, and then to fill in the rest of the circle by appropriate
exchanges and sign changes. The shuffling at the end to fill in all
four quadrants is not easy to do if multiple angles are being computed
at once in a vector processor, since each quadrant requires a different
shuffle.

\section{Methodology}

The approach taken in this work is slightly different than the approach
used historically. The goal of it is to avoid the relatively expensive
testing and shuffling resulting from the reduction of the angle to
one quadrant. To accomplish this, the angle is first scaled by $1/2\pi$
and reduced modulo $1$ onto $[-1/2, 1/2)$ by rounding and subtraction.
Then, a power series of the functions of an angle scaled down by a
power of two is computed, and the resulting trig pair is shifted back
onto the whole circle using double-angle formula recursion. 

Computing functions in this manner has a number of useful properties.
As discussed, it is highly vectorizable. It also produces a very smooth
result; that is, except for the error when the modular conversion
wraps around at $(2n+1)\pi$, there are no seams at which one must
eliminate discontinuites. Thus, algorithms which depend on differentiation
(\emph{e.g.} nonlinear least-squares fitting) will not be likely to
be disrupted by the joining. In reality, most IEEE-754 compliant algorithms
put a great deal of effort into avoiding such discontinuities by assuring
the error in the computation is less than 0.5~LSB. This effort, though
costs time. The method described here, instead, sacrifices a little
absolute accuracy in trade for extremely high speed, while preserving
smoothness.

In the implementation shown in this paper, I compute $\sin(2\pi x/4)$
and $\cos(2\pi x/4)$ where $x$ is the scaled and reduced angle as
discussed above, and then use 2 angle-doublings to expand this over
the entire circle using: 

\begin{eqnarray}
\sin(2\theta) & = & 2\sin\theta\cos\theta\label{eq:one}\\
\cos(2\theta) & = & \cos^{2}\theta-\sin^{2}\theta\nonumber \end{eqnarray}

The issue that needs to be addressed when carrying out any recursion
relation such as the repeated double-angle relations above is that
of stability. Since this algorithm is designed to generate single-precision
functions, and to operate on a single-precision vector processing
unit, one must pay close attention to instabilities in this recursion
due to roundoff error. In fact, for this algorithm, there were two
candidates for the angle doubling. The first is the one above, and
the second is\begin{eqnarray}
\sin(2\theta) & = & 2\sin\theta\cos\theta\label{eq:two}\\
\cos(2\theta) & = & 1-2\sin^{2}\theta\nonumber \end{eqnarray}

Each of the two methods has an instability, but in the chosen method
it is easily fixed at the end, and in the rejected method it is not
so.

Consider the case of a small error introduced: define \[
s_{1}=a\sin(\theta+\delta),\, c_{1}=a\cos(\theta+\delta)\]
 and iterate once using (\ref{eq:one}) to get:\begin{eqnarray*}
s_{2} & = & 2a^{2}\cos(\theta+\delta)\sin(\theta+\delta)=a^{2}\sin2(\theta+\delta)\\
c_{2} & = & a^{2}\cos^{2}(\theta+\delta)-a^{2}\sin^{2}(\theta+\delta)\\
 & = & a^{2}\cos2(\theta+\delta)\end{eqnarray*}
 Note that, then, the magnitude has drifted, but the error in the
angle is stable. 

Now, consider the case using (\ref{eq:two}):

\begin{eqnarray*}
s_{2} & = & a^{2}\sin2(\theta+\delta)\\
c_{2} & = & 1-2a^{2}\sin^{2}(\theta+\delta)\\
 & = & a^{2}\left(\cos^{2}(\theta+\delta)-\sin^{2}(\theta+\delta)\right)+(1-a^{2})\\
 & = & a^{2}\cos2(\theta+\delta)+(1-a^{2})\end{eqnarray*}
thus mixing the angle and amplitude error inextricably. 

Thus, the form of (\ref{eq:one}) produces a pair (after $n$ iterations)
of $a^{2n}\cos2^{n}(\theta+\delta)$, $a^{2n}\sin2^{n}(\theta+\delta)$
which has the same relative error in the angle as the initial estimate,
but a scale factor. Since the scale factor has no effect on the accuracy
of the properly-normalized trig pair, there are two options for computing
the raw pair. These options differ in the fitting algorithm used to
find the coefficients for the series.

The first method tries to produce a pair which has a very small normalization
error by directly computing least-squares series for sine and cosine.
Assuming the original amplitude error is small, so $a=1+\epsilon$,
then $a^{2n}\approx1+2n\epsilon$. Thus, the final amplitude has a
relative error $2n$ times bigger than the initial error. For $n=2$
and single-precision arithmetic, this implies an amplitude error of
about $4\times10^{-7}$ (assuming 1 LSB error at the start). To correct
for this at the end, compute $s_{n}^{2}+c_{n}^{2}=1+\alpha$ and divide
the function pair by $\sqrt{1+\alpha}$. In practice, since $\alpha$
is small, multiplying the pair by $1-\alpha/2=(3-s^{2}-c^{2})/2$
provides a correctly normalized result. Using this method, the RMS
error is $1.2\times10^{-7}$and the maximum error is $4.8\times10^{-7}$.
These errors are measured as $\sqrt{(\cos\theta-\textrm{ref\_}\cos\theta)^{2}+(\sin\theta-\textrm{ref\_}\sin\theta)^{2}}$
where the reference functions are computed to double precision using
the standard system library calls. The maximum amplitude error, computed
as $1-\sqrt{\cos^{2}\theta+\sin^{2}\theta}$ is $1.8\times10^{-7}$.
For this method, the polynomials:\begin{eqnarray*}
\sin\theta/4 & = & x(\,1.5707963235\\
 &  & -0.645963615\, x^{2}\\
 &  & +0.0796819754\, x^{4}\\
 &  & -0.0046075748\, x^{6})\\
\cos\theta/4 & = & 1\\
 &  & -1.2336977925\, x^{2}\\
 &  & +0.2536086171\, x^{4}\\
 &  & -0.0204391631\, x^{6}\end{eqnarray*}
are used, where $x$ is the range-reduced value of $\theta/2\pi$
discussed at the beginning of this paper.

The second method relaxes the requirement that the magnitude error
be small, by fitting the input angle to $\arctan(\sin\theta/\cos\theta)$,
but requires then an accurate division by the actual magnitude of
the pair at the end. Doing this produces a slightly better errror
budget, with an RMS error of $9.8\times10^{-8}$ and a maximum error
of $3.8\times10^{-7}$. The penalty in the final algorithm is doing
a precise reicprocal and multiply, which costs about 5\% in speed.
For this method, the polynomials:\begin{eqnarray*}
\sin^{\dagger}\theta/4 & = & x(\,1.5707963268\\
 &  & -0.6466386936\, x^{2}\\
 &  & +0.0679105987\, x^{4}\\
 &  & -0.0011573807\, x^{6})\\
\cos^{\dagger}\theta/4 & = & 1\\
 &  & -1.2341299769\, x^{2}\\
 &  & +0.2465220241\, x^{4}\\
 &  & -0.0123926179\, x^{6}\end{eqnarray*}
are used. Note that calling these sine and cosine is somewhat dangerous;
they are really the numerator and denominator of a good rational-function
representation for $\tan\theta/4$ but do not satisfy $\sin^{2}+\cos^{2}=1$.
I have annotated the terms in the equations with the ($\dagger$)
symbol to remind the reader of this.

\section{Results}

In figure~\ref{code-listing}, I show a sample implementation of
this, coded in C (compatible with gcc3.xx or IBM xlc). This code computes
a trig pair vector (4 trig pairs) in 140~ns on a 500~MHz PowerPC~7400
(aka PowerMacintosh G4). 

It is important to note that, even for this algorithm, the throughput
can be improved quite a bit. Because of the chain-computation nature
of this, there are many stalls waiting for latency in the vector unit.
If one computes two vectors at a time, by just interleaving this code
with itself with different variable names, throughput rises another
50\%. However, the logistics of using this in other code often does
not make it worthwhile. Also, the nature of optimizations of this
type is quite architecture-dependent and therefore not within the
intended scope of this work. However, with this optimization, one
can generate a single-precision trig function every 7 clock cycles
of the CPU (8 pairs in 100~ns), including loading and storing the
results. For a discussion of the instruction set and throughput of
this particular vector unit, see \cite{altivec}.

\section{Conclusions}

For applications requiring very high speed, random-access calculations
of sine-cosine pairs, it is possible to forego a very small amount
of accuracy in exchange for a great deal of speed. Using multiple-angle
formulas to expand a sine-cosine pair computed on a small piece of
the reduced range is an effective approach to this. The author has
been applying this method to the rapid computation of optical diffraction
patterns via Huygen's principle, but it should be widely applicable
to other problems for which very large numbers of single-precision
trigonometric pairs are needed.

\begin{biography}
{Marcus H. Mendenhall} is a Research Associate Professor of Physics
at Vanderbilt University. He received his Ph.D. from CalTech in 1983.
He has been working recently on pulsed, tunable Xray sources based
on Compton backscattering of laser photons from a relativistic electorn
beam. Techniques of numerical computation and analysis are a major
side-interest.

\begin{figure*}[p]
\begin{lyxcode}

{\scriptsize /{*}~define~FASTER\_SINCOS~for~the~slightly-less-accurate~results~in~slightly~less~time~{*}/}{\scriptsize \par}

{\scriptsize \#define~FASTER\_SINCOS}{\scriptsize \par}

~{\scriptsize ~~~~~}{\scriptsize \par}

{\scriptsize \#if~!defined(FASTER\_SINCOS)~/{*}~these~coefficients~generate~a~badly~un-normalized~sine-cosine~pair,~but~the~angle~is~more~accurate~{*}/}{\scriptsize \par}

{\scriptsize \#define~ss1~1.5707963268}{\scriptsize \par}

{\scriptsize \#define~ss2~-0.6466386396~}{\scriptsize \par}

{\scriptsize \#define~ss3~0.0679105987~}{\scriptsize \par}

{\scriptsize \#define~ss4~-0.0011573807}{\scriptsize \par}

{\scriptsize \#define~cc1~-1.2341299769~}{\scriptsize \par}

{\scriptsize \#define~cc2~~0.2465220241~}{\scriptsize \par}

{\scriptsize \#define~cc3~-0.0123926179}{\scriptsize \par}

{\scriptsize \#else~/{*}~use~20031003~coefficients~for~fast,~normalized~series{*}/}{\scriptsize \par}

{\scriptsize \#define~ss1~1.5707963235}{\scriptsize \par}

{\scriptsize \#define~ss2~-0.645963615~}{\scriptsize \par}

{\scriptsize \#define~ss3~0.0796819754~}{\scriptsize \par}

{\scriptsize \#define~ss4~-0.0046075748}{\scriptsize \par}

{\scriptsize \#define~cc1~-1.2336977925~}{\scriptsize \par}

{\scriptsize \#define~cc2~~0.2536086171~}{\scriptsize \par}

{\scriptsize \#define~cc3~-0.0204391631}{\scriptsize \par}

{\scriptsize \#endif}{\scriptsize \par}

{\scriptsize inline~void~FastSinCos(vector~float~v,~struct~phase~{*}ph)}{\scriptsize \par}

{\scriptsize \{}{\scriptsize \par}

~{\scriptsize ~~~~~~~vector~float~s1,~s2,~c1,~c2,~fixmag1;}{\scriptsize \par}

~{\scriptsize ~~~~~~~}{\scriptsize \par}

~{\scriptsize ~~~~~~~vector~float~x1=vec\_madd(v,~(vector~float)(1.0/(2.0{*}3.1415926536)),~}{\scriptsize \par}

~{\scriptsize ~~~~~~~~~~(vector~float)(0.0));}{\scriptsize \par}

~{\scriptsize ~~~~~~~}{\scriptsize \par}

~{\scriptsize ~~~~~~~/{*}~q1=x/2pi~reduced~onto~(-0.5,0.5),~q2=q1{*}{*}2~{*}/}{\scriptsize \par}

~{\scriptsize ~~~~~~~vector~float~q1=vec\_nmsub(vec\_round(x1),~(vector~float)(1.0),~x1);~}{\scriptsize \par}

~{\scriptsize ~~~~~~~vector~float~q2=vec\_madd(q1,~q1,~(vector~float)(0.0));}{\scriptsize \par}

~{\scriptsize ~~~~~~~}{\scriptsize \par}

~{\scriptsize ~~~~~~~s1=~~~~~vec\_madd(q1,~~~}{\scriptsize \par}

~{\scriptsize ~~~~~~~~~~~~~~~~~~~~~~~vec\_madd(q2,~}{\scriptsize \par}

~{\scriptsize ~~~~~~~~~~~~~~~~~~~~~~~~~~~~~~~vec\_madd(q2,~}{\scriptsize \par}

~{\scriptsize ~~~~~~~~~~~~~~~~~~~~~~~~~~~~~~~~~~~~~~~vec\_madd(q2,~(vector~float)(ss4),~}{\scriptsize \par}

~{\scriptsize ~~~~~~~~~~~~~~~~~~~~~~~~~~~~~~~~~~~~~~~(vector~float)(ss3)),~}{\scriptsize \par}

~{\scriptsize ~~~~~~~~~~~~~~~~~~~~~~~~~~~~~~~(vector~float)(~ss2)),~}{\scriptsize \par}

~{\scriptsize ~~~~~~~~~~~~~~~~~~~~~~~(vector~float)(ss1)),~}{\scriptsize \par}

~{\scriptsize ~~~~~~~~~~~~~~~(vector~float)(0.0));}{\scriptsize \par}

~{\scriptsize ~~~~~~~c1=~~~~~vec\_madd(q2,~}{\scriptsize \par}

~{\scriptsize ~~~~~~~~~~~~~~~~~~~~~~~vec\_madd(q2,~}{\scriptsize \par}

~{\scriptsize ~~~~~~~~~~~~~~~~~~~~~~~~~~~~~~~vec\_madd(q2,~(vector~float)(cc3),~}{\scriptsize \par}

~{\scriptsize ~~~~~~~~~~~~~~~~~~~~~~~~~~~~~~~(vector~float)(cc2)),~}{\scriptsize \par}

~{\scriptsize ~~~~~~~~~~~~~~~~~~~~~~~(vector~float)(cc1)),~}{\scriptsize \par}

~{\scriptsize ~~~~~~~~~~~~~~~(vector~float)(1.0));}{\scriptsize \par}

~{\scriptsize ~~~~~~~/{*}~now,~do~one~out~of~two~angle-doublings~to~get~sin~\&~cos~theta/2~{*}/}{\scriptsize \par}

~{\scriptsize ~~~~~~~c2=vec\_nmsub(s1,~s1,~vec\_madd(c1,~c1,~(vector~float)(0.0)));}{\scriptsize \par}

~{\scriptsize ~~~~~~~s2=vec\_madd((vector~float)(2.0),~vec\_madd(s1,~c1,~(vector~float)(0.0)),~}{\scriptsize \par}

~{\scriptsize ~~~~~~~~~~(vector~float)(0.0));}{\scriptsize \par}

~{\scriptsize ~~~~~~~/{*}~now,~cheat~on~the~correction~for~magnitude~drift...~}{\scriptsize \par}

~{\scriptsize ~~~~~~~~~~~~~~~if~the~pair~has~drifted~to~(1+e){*}(cos,~sin),~}{\scriptsize \par}

~{\scriptsize ~~~~~~~~~~~~~~~the~next~iteration~will~be~(1+e){*}{*}2{*}(cos,~sin)~}{\scriptsize \par}

~{\scriptsize ~~~~~~~~~~~~~~~which~is,~for~small~e,~(1+2e){*}(cos,sin).~~}{\scriptsize \par}

~{\scriptsize ~~~~~~~~~~~~~~~However,~on~the~(1+e)~error~iteration,~}{\scriptsize \par}

~{\scriptsize ~~~~~~~~~~~~~~~sin{*}{*}2+cos{*}{*}2=(1+e){*}{*}2=1+2e~also,~}{\scriptsize \par}

~{\scriptsize ~~~~~~~~~~~~~~~so~the~error~in~the~square~of~this~term}{\scriptsize \par}

~{\scriptsize ~~~~~~~~~~~~~~~will~be~exactly~the~error~in~the~magnitude~of~the~next~term.~~}{\scriptsize \par}

~{\scriptsize ~~~~~~~~~~~~~~~Then,~multiply~final~result~by~(1-e)~to~correct~{*}/}{\scriptsize \par}

~{\scriptsize ~~~~~~~~~~~~~~~}{\scriptsize \par}

{\scriptsize \#if~defined(FASTER\_SINCOS)~/{*}~this~works~with~properly~normalized~sine-cosine~functions,~but~un-normalized~is~more~accurate~overall~{*}/}{\scriptsize \par}

~{\scriptsize ~~~~~~~fixmag1=vec\_nmsub(s2,s2,~vec\_nmsub(c2,~c2,~(vector~float)(2.0)));}{\scriptsize \par}

{\scriptsize \#else~/{*}~must~use~this~method~with~un-normalized~series,~since~magnitude~error~is~large~{*}/}{\scriptsize \par}

~{\scriptsize ~~~~~~~fixmag1=Reciprocal(vec\_madd(s2,s2,vec\_madd(c2,c2,(vector~float)(0.0))));}{\scriptsize \par}

{\scriptsize \#endif}{\scriptsize \par}

~{\scriptsize ~~~~~~~c1=vec\_nmsub(s2,~s2,~vec\_madd(c2,~c2,~(vector~float)(0.0)));}{\scriptsize \par}

~{\scriptsize ~~~~~~~s1=vec\_madd((vector~float)(2.0),~vec\_madd(s2,~c2,~(vector~float)(0.0)),~}{\scriptsize \par}

~{\scriptsize ~~~~~~~~~~(vector~float)(0.0));}{\scriptsize \par}

~{\scriptsize ~~~~~~~}{\scriptsize \par}

~{\scriptsize ~~~~~~~ph->c=vec\_madd(c1,~fixmag1,~(vector~float)(0.0));}{\scriptsize \par}

~{\scriptsize ~~~~~~~ph->s=vec\_madd(s1,~fixmag1,~(vector~float)(0.0));}{\scriptsize \par}

{\scriptsize \}}{\scriptsize \par}

\end{lyxcode}

\caption{Sample Implementation for PowerPC AltiVec vector processor\label{code-listing}}
\end{figure*}
\end{biography}

\end{document}